\documentclass[preprint]{JHEP3} 

\usepackage{graphicx}
\usepackage{latexsym}

\def\be{\begin{equation}}
\def\ee{\end{equation}}
\def\ba{\begin{array}}
\def\ea{\end{array}}
\def\bea{\begin{eqnarray}}
\def\eea{\end{eqnarray}}
\def\bd{\begin{displaymath}}
\def\ed{\end{displaymath}}

\def\d{\delta}

\newcommand{\dr}{\raise.3ex\hbox{$\stackrel{\leftarrow}{\partial }$}{}}
\newcommand{\delr}{\raise.3ex\hbox{$\stackrel{\leftarrow}{\delta }$}{}}
\newfont{\twolineletters}{msbm10}

\csname @addtoreset\endcsname{equation}{section}

\newcommand\fverb{\setbox\pippobox=\hbox\bgroup\verb}
\newcommand\fverbdo{\egroup\medskip\noindent%
            \fbox{\unhbox\pippobox}\ }
\newcommand\fverbit{\egroup\item[\fbox{\unhbox\pippobox}]}
\newbox\pippobox

\title{Quantization of the Chern-Simons Coupling Constant}

\author{Xavier Bekaert \\ Dipartimento di Fisica, Universit\`a degli
Studi di Padova, \\ Via F. Marzolo 8, 35131 Padova, Italy\\
    E-mail: \email{bekaert@pd.infn.it}}

\author{Andr\'es Gomberoff \\ Centro de Estudios Cient\'{\i}ficos (CECS), Casilla 1469, Valdivia, Chile\\
    E-mail: \email{andres@cecs.cl}}

\preprint{DFPD/02/TH/27\\CECS-PHY-02/10\\
{\tt hep-th/0212099}}

\abstract{We investigate the quantum consistency of $p$--form
Maxwell--Chern--Simons electrodynamics in $3p+2$ spacetime
dimensions (for $p$ odd). These are the dimensions where the
Chern--Simons term is cubic, {\it i.e.}, of the form $F\wedge
F\wedge A$. For the theory to be consistent at the quantum level
in the presence of magnetic and electric sources, we find that the
Chern--Simons coupling constant must be quantized. We compare our
results with the bosonic sector of eleven dimensional supergravity
and find that the Chern--Simons coupling constant in that case
takes its corresponding minimal allowed value.}


\begin{document}

\section{Introduction}

\par Yang-Mills gauge theory in $2+1$ dimensions with a Chern--Simons (CS)
term (also called topological mass term) is known to have its
topological mass quantized for gauge groups with a non-trivial
third homotopy group \cite{Deser:1982}. The argument goes as
follows. The CS term added to the Yang-Mills action preserves
local gauge invariances for spacetime manifolds without boundary,
but is not invariant under large gauge transformations.  Under
large gauge transformations, the total action varies by a term
proportional to the instanton number representing the class of
$\pi_3(G)$. If we ask for the invariance of the path integral
under such topologically non-trivial gauge transformations, we get
a quantization condition of the CS coupling constant.

In general, the CS coupling constant is quantized since
$\pi_3(G)\simeq {\mbox{\twolineletters Z}}$ for any compact
connected simple Lie group $G$ \cite{Weinberg}. However, in the
Abelian case with compact group manifold $G=U(1)$ all the homotopy
groups higher than $\pi_1$ are trivial and, a priori, the CS
coupling constant can take arbitrary values. The authors of
\cite{Henneaux:1986} have pointed out that when electric and
magnetic charges are present the CS coupling constant must be
quantized just as in the non-Abelian case. The quantization arises
from two key features. The first, is that in the presence of
magnetic sources the electric charge is non-conserved for
non-vanishing CS coupling constant, hence electric worldlines may
end on magnetic sources. The value of the electric charge created
or annihilated on a magnetic charge is related to the CS coupling
constant. The second key property is the usual Dirac charge
quantization condition that was shown to remain valid in the
presence of a CS term. It  was also pointed out in
\cite{Henneaux:1986} that their result may be straightforwardly
generalized to $p$-form electrodynamics in $(2p+1)$-dimensional
spacetime, where the electric $(p-1)$-branes have the same
dimensionality as the Dirac brane attached to the magnetic
$(p-2)$-brane.

The quantization of the Abelian CS coupling constant was rederived
in \cite{Polychronakos:1987} for spacetimes with topology
$S^1\times M^2$ in the presence of a non-vanishing total magnetic
flux on $M^2$. Since the spacetime manifold contains a one-cycle
$S^1$, a quantization condition is expected from the fact that
$\pi_1\left(U(1)\right)\simeq {\mbox{\twolineletters Z}}$. Similar
phenomena occur at finite temperature \cite{Bralic:1996}. In this
case we are interested in Euclidean spacetimes, where the time
direction is effectively compactified into a circle.

All the analysis cited above considered odd-dimensional spacetimes
with linear equations of motion. The next step would be to
consider Abelian theories with a cubic CS term in the action. This
is the subject of the present paper. The non-Abelian case has been
considered in CS gravity, and it also leads to the quantization of
the CS coupling constant, which corresponds, in that case, to the
quantization of Newton's constant \cite{Zanelli}. For
$p$-form electrodynamics, a cubic term exists only in
$(3p+2)$-dimensional spacetime. Furthermore, if we want a
non-vanishing CS term $p$ has to be odd since the gauge group is
Abelian. The extended objects carrying electric charge in these
theories are $(p-1)$-branes. Magnetic charge is carried by
spacelike
$2p-1$--dimensional extended objects.

Cubic CS terms in  $p$-form electrodynamics appear, for instance,
in the bosonic sector of many supergravity theories. An example of
such a theory is the celebrated eleven-dimensional supergravity
\cite{Cremmer:1978}, which is currently believed to describe
the low energy effective action of M-theory.  Five-dimensional
supergravity also contains a CS term. This theory is known to
resemble $D=11$ supergravity in many respects \cite{Cremmer:1980}
and could be used as a toy model to test various ideas of
$M$-theory in a simpler setting (see also \cite{Boyarsky:2002} for
another five-dimensional toy model). This similarity arises from
the fact that
$D=5$ supergravity can be realized as a specific truncation of a
Calabi-Yau compactification of $D=11$ supergravity
\cite{Cadavid:1995}.

Because of its simplicity, the paper will consider only
five-dimensional Maxwell--Chern--Simons (MCS)
electrodynamics, which is the first example
of the theories we are looking for. This case is analogous to the
bosonic sector of $D=5$ simple supergravity in a fixed
gravitational background. The analysis will use differential forms
so that generalization to higher dimensions is straightforward.

\section{A 5D model of a gauge field with a Chern--Simons term}

MCS theory coupled to electric sources may be
described, in five-dimensional spacetime ${\cal M}_5$ by the
following action,
\begin{equation}
I = \frac{1}{2} \int F\wedge {}^*\!F + \frac{\alpha}{6}\int F
\wedge F \wedge A - \int A \wedge{}^*\!J_e \ ,\label{Ed}
\end{equation}
where $A$ is the gauge field, $F=dA$ its curvature, $J_e$ the
electric current and $\alpha$ a dimensionfull constant called the
CS coupling constant. The electric sources are pointlike objects,
which, by analogy with eleven-dimensional supergravity, we may
call them M$0$-branes. The equations of motion and Bianchi
identity are
\begin{eqnarray}
d{}^*\! F +\frac{\alpha}{2} F\wedge F &=& {}^*\!J_e \label{24}\\
dF &=& 0\, \label{25} .
\end{eqnarray}
The action is gauge invariant if and only if $J_e$ is a conserved
current, {\em i.e.}, if  $d^*\!J_e=0$. This requirement is
consistent with the equations of motion, as can be seen by taking
the exterior derivative of (\ref{24}), which gives precisely
(\ref{25}). We may take the electric current to be generated by
the 1--dimensional worldvolume, ${\cal M}_1$, of a particle of
charge
$e$, parameterized by $z^{\mu}(\tau)$. The dual of its associated
current may be written as ${}^*\!J_e=e\,P({\cal M}_1)$, where the
four--form $P({\cal M}_1)$ is the Poincar\'e dual of ${\cal M}_1$.
(In Appendix A we introduce  the definition and some key
properties of Poincar\'e duality. For more details on this subject
and its use in electrodynamics see \cite{Lechner:2001}.)

Now we introduce magnetic sources.  In 5D,
${}^*\!F$ is a 3--form, and therefore these will correspond to
one-dimensional extended objects: ``magnetic strings'', or
M$1$-branes. The magnetic string current is described by a
2--form, $J_m$. The field  equations now  take the form
\begin{eqnarray}
d{}^*\! F +\frac{\alpha}{2} F\wedge F &=& {}^*\!J_e  \label{m1} \\
dF &=& {}^*\!J_m  \label{m2} \, .
\end{eqnarray}
These equations are gauge invariant, but the electric current
$J_e$ is not, in general,  conserved. In fact, taking the exterior
derivative of (\ref{m1}) gives
\begin{equation}
d{}^*\!J_e =\alpha {}^*\!J_m \wedge F \, .\label{nc}
\end{equation}
This last equation is telling us that electric current does not
need to be conserved on the worldsheet of a magnetic string,
${\cal M}_2$, unless the flux of the  field strength across ${\cal M}_2$
is required to vanish. Since the magnetic current $J_m$ is
conserved, its worldsheet, ${\cal M}_2$, is a two-dimensional
surface with no boundaries. The dual of the magnetic charge may be
written:
${}^*\!J_m=g\,P({\cal M}_2)$, where $g$ is a constant measuring
the charge of the magnetic source.

Note  that the present situation is different from both the
standard Maxwell case and the quadratic Chern--Simons studied in
\cite{Henneaux:1986}. Indeed, Eq. (\ref{nc}) shows that we cannot
specify external sources, because the conservation rule for $J_e$
depends explicitly on the field strength $F$. This is a major
obstacle in the construction of a variational principle giving
rise to Eqs. (\ref{m1}) and (\ref{m2}). One solution would be to
consider the manifold ${\cal M}_2$ to be a surface where boundary
conditions must be prescribed. However, here we adopt a different
strategy, inspired by eleven--dimensional supergravity. We will
add degrees of freedom living on the magnetic string worldsheet.
These new degrees of freedom will dynamically introduce the
``boundary conditions" for
$F$. In the next section we will present an action principle
giving rise to equations (\ref{m1}) and (\ref{m2}).

\section{The Action}

Consider the modified Bianchi identity (\ref{m2}).  Following the
standard procedure, we introduce a Dirac worldvolume,  ${\cal
N}_3$, whose boundary is the magnetic string worldvolume, ${\cal
M}_2=\partial{\cal N}_3$, and solve it by writing the field
strength as $F= dA + ^*\!G$. Here  $^*\!G=P({\cal N}_3)$, and
therefore
$^*\!J_m=d^*\!G$ (see Appendix A). On the magnetic  string worldvolume, ${\cal
M}_2$, we shall also introduce a scalar field $\Phi$, which
describes the new degrees of freedom discussed in the previous
section. Consider now the following action principle  in 5
spacetime dimensions:
\begin{eqnarray}
I &=& \frac{1}{2}\int {}^*\!F\wedge F + \frac{\alpha}{6}\int dA
\wedge dA \wedge A
+\frac{\alpha}{2}\int {}^*\! G\wedge A \wedge dA \nonumber \\
& &- \int A \wedge {}^*\!J_e  +  \frac{\alpha}{2} \int \left(
-f\wedge A + \frac{1}{2} {}^{{*}}\!f \wedge f + \omega \Phi
{}^{{*}}\!j
\right)\wedge {}^*\!J_m
   +  I_k\,.
\label{I}
\end{eqnarray}
Here
\begin{equation}
f= d\Phi - A\mid_{pullback} +{{}^*\!}g_{-1} \ ,
\label{f}
\end{equation}
where $A\mid_{pullback}$ denotes the pullback of $A$ on ${\cal
M}_2$, and ${{}^*\!}$ is the Hodge star on the worldvolume ${\cal
M}_2$. From now on, we will omit the subscript ``pullback" which
should be self-evident from the context. The parameter  $\omega$
is any real number and $I_k$ are kinetic terms. Finally, $g_{-1}$
defines a set of Dirac worldlines that originate instantons
(localized at one spacetime event) on the magnetic string
worldsheet. The case of one instanton of strength $\nu$ located at
the point ${\cal M}_0$ (denoted in that way to make higher
dimensional generalization straightforward) is described by the
dyonic instanton ``current" ${{}^*\!}j=\nu\,P({\cal M}_0)$, a
$0$-form proportional to the Poincar\'e dual to ${\cal M}_0$ in
${\cal M}_2$. The Dirac point is defined by a worldline ${\cal
N}_1$ ending or originating at the instanton, in such a way that
${}^*\! j= d{{}^*\!}g$ if ${{}^*\!}g=\nu\,P({\cal N}_1)$.

\begin{figure}[ht!]
     \centerline{\includegraphics[width=0.3\linewidth]{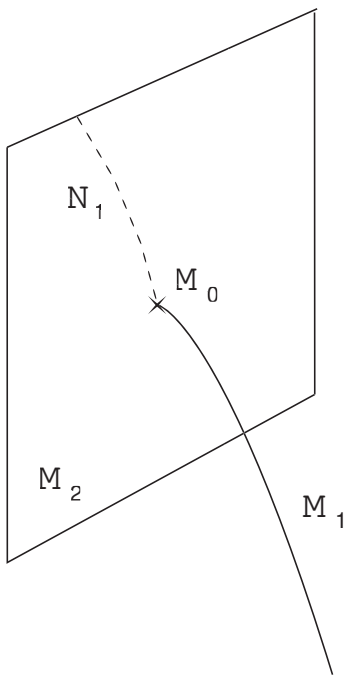}}
\caption{An electric worldline, ${\cal M}_1$, ends on a magnetic worldsheet ${\cal M}_2$ and
produces a magnetic instanton, ${\cal M}_0$.}
\end{figure}

Note that, for simplicity of notation, sometimes we make use of
$d\Phi$ and ${{}^*\!}g_{-1}$ in the action as $5$-dimensional forms.
These should be understood as arbitrary extensions whose pullback
gives the corresponding $2$--dimensional differential form on the
string worldsheet.

\section{Gauge invariances}

The action (\ref{I}) is invariant under two different gauge
transformations.  First,
\begin{eqnarray}
A &\longrightarrow& A + d\Lambda \label{gt1} \\
\Phi &\longrightarrow& \Phi + \Lambda \label{gt2} ,
\end{eqnarray}
where, in the second equation, $\Lambda$ is understood as the
pullback along the string. In the absence of magnetic sources,
this transformation reduces to the standard gauge transformation
of electrodynamics. In the present case, the invariance of the
action under (\ref{gt1}) and (\ref{gt2}) requires that the
following identity is satisfied:
\begin{equation}
d{}^*\!J_e = \frac{\alpha}{2}(1-\omega) \, {}^*\!j \wedge {}^*
\!J_m \ .
\label{nc2}
\end{equation}
This expression shows that electric charge does not need to be
conserved if $\omega \neq 1$. In fact, an electric worldline may
end on a magnetic worldsheet and create a dyonic instanton on it.
The eleven--dimensional analog is the well known fact that
M$2$-branes can end on M$5$-brane with a dyonic string as
intersection. In that case, a spacelike picture is possible
because there is ``more room" in eleven dimensions. Also note that
in the absence of instantons on the worldsheet of the string, we
get conservation of the electric charge for any value of $\omega$.
The Poincar\'e dual translation of the relation (\ref{nc2}) is
$\partial {\cal M}_1={\cal M}_0\subset {\cal M}_2$ since
${{}^*\!}j \wedge {}^*\!J_m$ is proportional to  the Poincar\'e
dual of ${\cal N}_1$ in ${\cal M}_5$ (${}^*\!J_m$ first projects
on ${\cal M}_2$ and then ${{}^*\!}j$ projects on ${\cal N}_1$ in
${\cal M}_2$).

The second gauge freedom of this system is  associated with the
position of the Dirac brane and can be described as follows.  We
may deform  ${\cal N}_3$ into a new manifold
${\cal N}_3^{\prime}$ sharing the same boundary ${\cal M}_2$. This new manifold is
equally appropriate for solving (\ref{m2}) in the way indicated
above.  The field strength $F$ is invariant under this
displacement of the Dirac brane, but
\begin{eqnarray}
{}^*\!G^{\prime} & = &{}^*\!G+d^*\!V \ ,  \\
A^{\prime}  &=& A - {}^*\!V \  .\label{varA}
\end{eqnarray}
Here, $^*\!V =g P({\cal V})$ is the one-form dual to the manifold
${\cal V}$ swept while moving the Dirac worldvolume from ${\cal
N}_3$ to ${\cal N'}_3$, that is, $\partial {\cal V}= {\cal
N}_{3}^{\prime}-{\cal N}_3$. The field $f$ on the magnetic string
is also invariant under (\ref{varA}) because the pullback of
${}^*\!V$ vanishes. The action (\ref{I}) is invariant under small
displacements of the Dirac brane, that is, displacements such that
$\cal V$ does not intersect the worldvolume of any other object.
This shows that the Dirac membrane is unobservable.

The same remark applies to the Dirac point living on the magnetic
string. The corresponding gauge transformation is
\begin{eqnarray}
{{}^*\!}g\rightarrow {{}^*\!}g + d{{}^*\!}v\,,\, \Phi \rightarrow
\Phi-{{}^*\!}v.\label{worldsheet}
\end{eqnarray}
The scalar ${{}^*\!}v$ is proportional to the Poincar\'e dual of
the surface described by the Dirac worldline in the string
worldsheet. The action is obviously invariant under
(\ref{worldsheet}).

\section{Equations of motion}

The equation of motion coming from varying $A$ is
\begin{equation}
d{}^*\!F +\frac{\alpha}{2} F\wedge F -
\frac{\alpha}{2}{}^*\!J_m\wedge ({}^*\! f + f)  = {}^*\!J_e \ .
\label{eom}
\end{equation}
Note that on deriving this equation we have made use of the
following identity, which is discussed in Appendix B,
\begin{equation}
{}^*\! G \wedge {}^*\!G = 0 \ \ .\label{singular}
\end{equation}
If conveniently regularized, this identity holds at any point
where the Dirac worldvolumes do not intersect, which will be the
case in general, because, as we shall see below, consistency of
the variational principle will require that Dirac membranes never
intersect themselves (Dirac veto).

Varying $\Phi$ we obtain,
\begin{equation}
\left(dA  + d^*\! f + \omega {}^*\! j \right) \wedge  {}^*\! J_m=0
\ . \label{eomphi}
\end{equation}
As a consistency check, note that taking the exterior derivative
of (\ref{eom}) and using (\ref{eomphi}) we get (\ref{nc2}). From
the definition of $f$, we also have, on ${\cal M}_2$, the Bianchi
identity
\begin{equation}
df = -dA + {}^*\! j \ \ .
\label{bianchif}
\end{equation}

Since the magnetic worldsheet is a Minkowskian two--dimensional
manifold, the 1--form field strength $f$ defined on it can be
decomposed into its self--dual part $f_+$ and anti--self--dual
part $f_-$,
\begin{equation}
f=f_+ + f_- \  ,  \quad  {}^*\!f_{\pm}=\pm f_\pm \ \ .
\label{fdec}
\end{equation}
Equation (\ref{eom})  can then be rewritten,
\begin{equation}
d{}^*\!F +\frac{\alpha}{2} F\wedge F - \alpha {}^*\!J_m\wedge  f_+
= {}^*\!J_e \ ,
\label{eq}
\end{equation}
while expressions (\ref{eomphi}) and (\ref{bianchif}) are
equivalent to
\begin{eqnarray}
df_+  + dA &=&  \frac{1-\omega}{2} \  {}^*\! j \ ,  \label{df+}\\
df_-    &=& \frac{1+\omega}{2} \ {}^*\! j  \ .
\end{eqnarray}
These last two equations are defined on the magnetic worldsheet.
We notice that the anti--self--dual part is decoupled from the
bulk field $A$, since $f_-$ does not appear in (\ref{eq}) nor in
(\ref{df+}).

When $\omega=-1$, it is therefore natural to consider the sector
of the theory where $f_-=0$, in which case the dyonic instanton is
self--dual. Furthermore, the equations (\ref{eq}) and (\ref{df+})
then describe the bosonic sector of simple D=5 supergravity
\cite{Cremmer:1980}. The worldsheet self-dual boson is precisely
the analog of the dynamics of the M$5$-brane in
eleven--dimensional supergravity. Strictly speaking, for these
theories, the self-duality condition must arise from the
variational principle. Indeed, it is possible to write an action
similar to (\ref{I}) for which $f_-=0$ arises as a consequence of
the variation of the action, using the PST technique
\cite{Sorokin:1997}. A regularized action has been proposed
recently in \cite{Lechner:20012}. Concerning the issue of
regularization in MCS theories with cubic CS term, one may look
also at \cite{Boyarsky:2002}.

Note however that one may consider other sectors of the theory for
arbitrary values of $\omega$. For instance, the sector $f_+=0$
leads to the equation (\ref{m1}) which does not contain explicitly
the magnetic membrane fields. Still, the presence of the magnetic
source imposes boundary conditions on $F$ via equation
(\ref{df+}). In particular, when $\omega=+1$, on the brane, the
electric field is orthogonal to the magnetic string.

\section{Charge Conservation}

It is a well known fact that the ``brane source charge" is not
conserved in theories with CS terms (for an extensive discussion
on this, see \cite{Don}). In the present case, as we have already
noticed, the electric source charge current, $J_e$ is not
conserved. Nevertheless, we know that the action (\ref{I}) is
gauge invariant in the standard way. We therefore expect a
conserved charge associated to this symmetry. We may discover this
conservation law from the equations of motion. In fact, from
(\ref{nc2}) we see that
\begin{equation}
d\left({}^*\!J_e-\frac{\alpha}{2}(1-\omega)\, {}^*\! g \wedge
{}^*\!J_m\right)=0\ .
\label{cons}
\end{equation}
This can be rephrased by saying that the Dirac worldline is the
geometric continuation of the electric worldline in the magnetic
string worldsheet.  The electric charge is therefore
``transferred" into the Dirac point. This is exactly the same
phenomena that was described previously in three spacetime
dimensions \cite{Henneaux:1986}.

Now we can define the following conserved charge,
\begin{equation}
Q=\int\limits_{D^4}({}^*\!J_e-\frac{\alpha}{2}(1-\omega)\,
{{}^*\!}g \wedge {}^*\!J_m)
\end{equation}
Here $D^4$ is a four--dimensional, spacelike ball, whose boundary
$\partial D^4=S^3$. is a three--sphere. Note that although $Q$ is
a conserved quantity in virtue of (\ref{cons}), it is not gauge
invariant. In fact, we can always deform the Dirac worldvolume so
that, for instance,  if initially it crossed $D^4$, at the end it
will no longer cross it. To define a gauge invariant charge, we
must define $Q$ in the limit where the radius of $S^3$ goes to
infinity.

Let us now take two different representatives, $D^4_{(1)}$ and
$D^4_{(2)}$. The first one does not intersect the magnetic
worldsheet while the second one has no intersection with the
worldline. If we evaluate the tension with $D^4_{(1)}$ we get
\begin{eqnarray}
\int\limits_{D^4_{(1)}}{}^*\!J_e=e\nonumber
\end{eqnarray}
and with $D^4_{(2)}$
\begin{eqnarray}
\frac{\alpha}{2}(1-\omega)\int\limits_{D^4_{(2)}}{{}^*\!}g \wedge {}^*\!J_m&=&\frac{\alpha
g}{2}(1-\omega)\int\limits_{S^1}{{}^*\!}g =\frac{\alpha
g}{2}(1-\omega)\int\limits_{B^2}{{}^*\!}j\nonumber\\&=&\frac{\alpha
g\nu }{2}(1-\omega)\nonumber
\end{eqnarray}
where $D^4_{(2)} \cap {\cal M}_2 = S^1 =\partial B^2$. In
conclusion, the electric charge $e$, created (or annihilated) on a
instanton of charge $\nu$, located on a magnetic string of charge
$g$ is given by
\begin{equation}
e=\frac{\alpha g\nu }{2}(1-\omega)\,.\label{charges}
\end{equation}
Note now that the Bianchi identity (\ref{df+}) implies that
\begin{equation}
f_+=d\Phi_+ -A +\frac{1-\omega}{2} \  {}^*\! g \,.
\end{equation}
This last equation combined with (\ref{eq}) leads to
\begin{equation}
d({}^*\!F+\frac{\alpha}{2}A\wedge dA+\alpha A\wedge {}^*\!G+\alpha\Phi_+ \\
{}^*\!J_m)={}^*\!J_e +\frac{\alpha}{2}(1-\omega){}^*\!J_m\wedge
{}^*\!g\,.
\end{equation}
from where we can rewrite $Q$ in terms of quantities defined at
spatial infinity only,
\begin{equation}
Q=\int_{S_{\infty}^3} {}^*\!F+\frac{\alpha}{2}A\wedge dA -
g\alpha\int_{\gamma}A -\alpha g \left(\Phi(+\infty)-\Phi(-\infty)
\right) \ .
\label{Q}
\end{equation}
Here $\gamma$ is the line where the Dirac membrane intersects the
sphere $S_{\infty}^3$. The quantities $\Phi(\pm \infty)$ are the
values of the scalar field $\Phi$ at the points where the magnetic
string intersects  $S_{\infty}^3$, where the sign in $\pm\infty$
is defined by the orientation of the string.
\section{Dirac vetos}

Let us now vary the action with respect to the Dirac worldvolume.
If we first fix its boundary (the string worldsheet) we obtain,
\begin{equation}
d{}^*\!F + \frac{\alpha}{2}dA \wedge dA\left|_{\mbox{\tiny
pullback on Dirac worldvolume}}\right.=0 \ .\label{Dirvet}
\end{equation}
For consistency with (\ref{eq}) this implies three Dirac vetos
\begin{itemize}
\item { \em No electric charge can be located on the Dirac worldvolume.} This is the standard Dirac veto.
\item { \em No magnetic charge can be located on the Dirac worldvolume.}
\item {\em Dirac worldvolumes cannot intersect each other (or themselves).} In other words, the pullback of ${}^*\! G$ on the Dirac worldvolume must vanish, so that $dA$ may be replaced by $F$ in (\ref{Dirvet}).
Note that the Dirac vetos, in the presence of the CS term, are
more restrictive than in the Maxwell case where only the first one
applies.

\end{itemize}
From the variation of the magnetic string we obtain the Lorentz
force equation acting on the magnetic strings. This equation will
depend on the precise form of the kinetic terms $I_k$ present in
the action (\ref{I}).

If we vary the action with respect to the Dirac wordline with the
instanton position fixed, we get
\begin{eqnarray}
d({{}^*\!}f + A)\left|_{\mbox{\tiny pullback on Dirac
worldine}}\right.= 0,
\end{eqnarray}
which implies, for consistency with (\ref{eomphi}), a Dirac veto
on the string worldsheet:
\begin{itemize}
\item{ \em No dyonic charge can be located on the Dirac worldine.} This is only necessary if $\omega\neq 0$.
\end{itemize}

\section{Quantization conditions}

\subsection{Quantization condition in the bulk}

Consider a finite displacement of a given Dirac membrane,
described, as in section 4,  by the manifold $\cal{V}$ swept by
the Dirac worldvolume as it is moved from ${\cal N}_3$ to ${\cal
N}_3^{\prime}$, so that
$\partial {\cal V}= {\cal N}_{3}^{\prime}-{\cal N}_3$. We again define
the Poincar\'e dual of this manifold, $^*\!V = P({\cal V})$. The
variation of the action is given by
\begin{equation}
\delta_V I = \int {}^*\!V\wedge {}^*\!J_e  \  .
\label{dI}
\end{equation}
Note that the terms coming from integrating over Dirac membranes,
Dirac points, magnetic strings or instantons vanish. This is due
to the Dirac vetos. In fact, $\cal{V}$ is a 4--dimensional,
compact, manifold. If there were, say, a magnetic string
worldsheet, ${\cal M}_2$, inside it (a non--compact 2--dimensional
manifold), then an intersection between ${\cal M}_2$ and $\partial
{\cal V}$ is unavoidable. But $\partial {\cal V}$ describes the
initial and final configurations of the magnetic string, and
therefore the intersection implies that either the final or the
initial configuration (or both) cannot satisfy the Dirac veto. The
same phenomena occurs for Dirac membranes. Now we can easily
integrate (\ref{dI}) to get,
\begin{equation}
\delta_V I = eg k \ \ ,
\label{q1}
\end{equation}
where $k \in{\mbox{\twolineletters Z}}$ is the number of electric
worldlines crossing through $\cal V$. Obviously, for an
infinitesimal variation we may always keep $\cal V$ free of
electric intersections, hence this anomaly appears only for finite
gauge transformations.

If we require the path integral to be invariant under (\ref{q1}),
we obtain the usual Dirac quantization condition
\begin{eqnarray}
eg &=& 2\pi n \label{aa}
\end{eqnarray}
for any integer $n$ (we have set $\hbar =1$).

\subsection{Quantization condition on the worldsheet}

In the case of the instanton the computation of the Dirac anomaly
is subtle, among other things because the instanton is dyonic. The
Dirac anomaly is,
\begin{eqnarray}
\delta_{v} I &=& \frac{\alpha g\omega}{2}\int\limits_{{\cal M}_2} {}^*\!v\, {{}^*\!}j\,.
\label{q2}
\end{eqnarray}
where $^*\!v = \nu P({\cal V}_2)$, and ${\cal V}_2$ is defined, on
the magnetic string worldsheet, by the variation of the Dirac
point worldline. Taking into account the fact that the instanton
is dyonic (and therefore lives on ${\cal V}_2$), it can be shown
that consistency at the quantum level imposes
\begin{equation}
\alpha g \nu^2\omega=2\pi m\label{aaa}
\end{equation}
with $m$ an integer. The quantization of an instanton in two
dimensions was previously considered in \cite{Henneaux:1985}. As
one can see, there is a subtle factor of two with respect to the
``naive" application of Dirac quantization condition with equal
electric and magnetic charge. This feature is generic for dyons
\cite{Deser:1997} (in dimensions 2 mod 4), and comes from the fact that the
phase $e^{-iI}$ must be strictly invariant only under gauge
transformations connected to the identity. The simplest system for
which this consideration applies contains two dyons, which multiplies
by two the total flux (for details see \cite{Deser:1997}).

\subsection{Quantization of the CS coupling constant}

From (\ref{charges}), (\ref{aa}) and (\ref{aaa}) we obtain
\begin{equation}
\alpha = \frac{\omega}{(1-\omega)^2}\frac{(e_0)^3}{\pi^2}\,N\,,
\quad\quad (N \in{\mbox{\twolineletters Z}}), \label{quant}
\end{equation}
which is the announced quantization condition of the CS coupling
constant $\alpha$, where $e_0$ is the minimal electric charge of
the MCS theory. The integer $N$ is directly related to the
(quantized) electric charge created by the minimal instanton
charge ($m=1$) living on the string with minimal magnetic charge
$2\pi\ e_0$. This formula is only valid when $\omega\neq 0,1$.
When $\omega=0$ the right hand side of (\ref{q2}) identically
vanishes, and therefore the quantization condition in the
worldsheet, (\ref{aaa}), is not required. The parameter $\alpha$
can therefore take any value in this case. When $\omega=1$, the
electric charge is conserved, and therefore we cannot relate,
through (\ref{charges}), the value of electric and magnetic
sources. Again, in this case, $\alpha$ is arbitrary.

When $\omega=-1$ and $f_-=0$, the system is (on-shell) the
5-dimensional analog of the MCS sector of 11-dimensional
supergravity coupled to M2-- and M5--branes. The coupling constant
$\alpha$, in that case, is fixed by the requirement of
supersymmetry, and it turns out to be given by the choice $N=1$ in
(\ref{quant}).

\section{Conclusions}

In the present paper we first constructed a gauge invariant action
principle which generalizes the usual MCS theory (for cubic CS
terms) so that it can be coupled to magnetic sources. It turns out
that there is no way to implement that by attaching Dirac branes
to the magnetic sources only. It is necessary -- as it is the case
when eleven--dimensional supergravity is coupled to M5--branes --
to add further degrees of freedom living on the magnetic source.
Then we studied the quantum consistency of the different possible
theories (parameterized by a real parameter, $\omega$), and
conclude that the CS coupling constant must be quantized according
to (\ref{quant}). Although we have used the 5--dimensional case
throughout this paper, the generalization for higher ($5$ modulo
$6$) dimensional spacetime is straightforward. One only needs to
substitute the different form fields by their analogs of higher
degree.

Bachas \cite{Bachas} obtained previously a similar quantization
condition\footnote{Compare the relation (\ref{quantB}) with
equation (4.22) of \cite{Bachas}. The different normalizations can
be translated into $\alpha=\sqrt{2} k\kappa_{(5)}$ and
$e_0=\sqrt{2}\kappa_{(5)}q$.}
\begin{equation}
\alpha = \frac{(e_0)^3}{(2\pi)^2}\,M\,,  \quad\quad (M
\in{\mbox{\twolineletters Z}})\,. \label{quantB}
\end{equation} His derivation was based on a different
argumentation, using compactification to four dimensions together
with the Witten effect. The two quantization conditions
(\ref{quant}) and (\ref{quantB}) coincide for $\omega=-1$.

Condition (\ref{quantB}) was obtained even earlier under the
assumption that $F$ was a non-trivial cocycle with integer periods
\cite{Duff:1995}, that is, in the absence of magnetic sources.
Indeed, since the five-dimensional spacetime manifold ${\cal M}_5$
is closed, it can be taken as the boundary of a six-dimensional
manifold ${\cal M}_6$. Therefore the CS term $F\wedge F\wedge A$
is lifted to $F\wedge F\wedge F$. Furthermore, physics should not
depend on the choice of the auxiliary manifold ${\cal M}_6$. Under
the assumption that the change $\d {\cal M}_6$ of six-dimensional
manifold is the direct product of three two-cycles on which $F$
takes integer periods, the condition (\ref{quantB}) follows from
the invariance of the path integral.

However, as pointed out by Witten \cite{Witten:1996}, a CS term is
quantum-mechanically inconsistent when $M$ is not a multiple of
$6$ because the integral of $F\wedge F\wedge F$ is not a multiple
of six for arbitrary closed manifolds $\d {\cal M}_6$. In other
words, the insertion of the action (\ref{Ed}) in the path integral
may lead to inconsistencies if the $U(1)$ bundle of the MCS theory
is non-trivial. Quantum consistency is restored by adding the
gravitational and fermionic sector of eleven-dimensional
supergravity, as shown in \cite{Witten:1996}, by taking into
account gravitational corrections plus a rather subtle argument
using $E_8$ gauge theory (see also \cite{Diaconescu:2000} for
subsequent developments).

We stress that, in contrast, we derived here (\ref{quant}) from
the presence of magnetic sources (thus $F$ is \textit{not} closed)
but with a \textit{trivial} $U(1)$ bundle and a topologically
\textit{trivial} spacetime manifold (i.e. ${\cal M}_5
\simeq\mathbb R^5$), thereby avoiding the above-mentioned quantum
inconsistency. Surprisingly enough, we still obtain the
supergravity factor in this extremely simple situation.
\\

{\bf Note added in proof} : While finishing the present article,
the work \cite{Kalkkinen:2002} appeared, which consider similar
issues regarding the construction of the action principle.  This
is done from a different perspective, and in the framework of
$M$--theory .

\section*{Acknowledgments}

The authors acknowledge K. Lechner, P. Marchetti, D. Marolf, D.
Sorokin, C. Teitelboim, M. Tonin, E. Witten, and specially Marc
Henneaux for useful discussions.

X.B. gratefully acknowledges support by the European Commission
TMR program HPRN-CT-00131. Institutional support to the Centro de
Estudios Científicos (CECS) from Empresas CMPC is gratefully
acknowledged. CECS is a Millennium Science Institute and is funded
in part by grants from Fundaci\'on Andes and the Tinker
Foundation. A.G. gratefully acknowledges support from FONDECYT
grants 1010449, 1010446 and from Fundaci\'on Andes. A.G. also
wishes to thank the kind hospitality of the Service de Physique
Th\'eorique et Math\'ematique at Universit\'e Libre de Bruxelles,
where part of this work was done.

\noindent
\appendix

\section{Review of Poincar\'e duality}\label{PD}

Let ${\cal M}_p$ be a smooth, oriented manifold of dimension $p$
embedded on a  $D$--dimensional manifold ${\cal M}$. Let
$x^{\mu}$, $\mu=1\ldots D$ be coordinates on ${\cal M}$. We
parameterize ${\cal M}_p$  with $p$ coordinates $\sigma^i$ by
$x^{\mu}=z^{\mu}(\sigma^i)$ and define the following $p$--tensor  on ${\cal M}$
\begin{equation}
V^{\mu_1 \cdots \mu_p}(x)= \int_{{\cal M}_p} \delta^{(D)}(x-z)
dz^{\mu_1} \cdots dz^{\mu_p} \  .
\label{a1}
\end{equation}
The Poincar\'e dual of ${\cal M}_p$, $P({\cal M}_p)$,  is the
$(D-p)$--form defined by
\begin{equation}
P({\cal M}_p) = {}^*\! V \ .
\label{poincare}
\end{equation}
The key, defining  property of the Poincar\'e dual is that  given
any $p$--form on ${\cal M}$, $\Omega$, then
\begin{equation}
\int\limits_{{\cal M}_p} \Omega\mid_{\tiny \mbox{pullback}}=
-\int\limits_{\cal M}P({\cal M}_p)\wedge\Omega  \label{a2}
\end{equation}
where $\Omega\mid_{\tiny \mbox{pullback}}$ is the pullback of
$\Omega$ on ${\cal M}_p$. Another important property of the
Poincar\'e dual is the  following:  If $\partial {\cal M}_p$ is
the boundary of ${\cal M}_p$, then
\begin{equation}
P(\partial {\cal M}_p) = (-)^{D-p+1}d P({\cal M}_p) \ . \label{a3}
\end{equation}
Both (\ref{a2}) and (\ref{a3}) can be derived straightforwardly
from (\ref{a1}), (\ref{poincare}).

\section{Regularization of ${}^*\!G\wedge{}^*\!G$}

In this appendix we show that the singular expression
\begin{equation}
{}^*\!G(x)\wedge{}^*\!G(x) \ \ ,
\label{sing}
\end{equation}
may be set locally to zero by  properly regularizing the delta
functions present on it. The idea is to give some small transverse
width
$\epsilon$ to the Dirac worldvolume, and then take the limit
$\epsilon\rightarrow 0$. For this purpose, we will consider, in a
neighborhood of the point $x$, a local set of coordinates on
spacetime such that the Dirac worldvolume is parallel to  the
coordinates $(x^0,x^1,x^2)$. The dual of expression (\ref{sing})
is a 1--form whose components are proportional to
\begin{equation}
\epsilon_{\mu\nu\rho\sigma\tau}\int \delta^{(5)}(x-y) dy^\alpha \wedge dy^\mu \wedge dy^\nu
\int \delta^{(5)}(x-\tilde{y}) d\tilde{y}^\rho \wedge d\tilde{y}^\sigma \wedge d\tilde{y}^\tau \ \ .
\label{du}
\end{equation}
Here both integrals are taken over the worldvolume of the Dirac
membrane, which may be parameterized by $\sigma^a$ ($a=0,1,2$),
setting $y^\mu\equiv y^\mu(\sigma^a)$. In the local set of
coordinates we chose, we can write $y^a=\sigma^a$, $y^3=y^4=0$.
The transversal width of the delta function is implemented by
taking
\begin{equation}
\delta^{(5)}(x-y(\sigma)) \longrightarrow \delta(x^a-\sigma^a)\Delta_\epsilon(x^3)\Delta_\epsilon(x^4) \ \ ,
\label{delta}
\end{equation}
where $\Delta_\epsilon$ is any regular function which is equal to
the delta function as $\epsilon$ vanishes. The deltas in the
longitudinal directions are kept unchanged, and can be integrated
so that expression (\ref{du}) is proportional to
\begin{equation}
  \epsilon_{abcde}\left(\Delta(x^3)\Delta(x^4)\right)^2
  \tilde{\epsilon}^{\alpha ab}\tilde{\epsilon}^{cde}  \ \ ,
\label{du2}
\end{equation}
where $\tilde{\epsilon}^{abc}$  is the Levi--Civita symbol of the
Dirac 3--dimensional worldvolume. It is clear that this quantity
is zero for any finite value of $\epsilon$, and in particular in
the limit $\epsilon\rightarrow 0$. Let us mention that the
previous argument can also be applied to any product of Poincar\'e
duals taken at the same point. Indeed, any such kinds of
identities (e.g. ${}^*\!V(x)\wedge{}^*\!V(x)$) have been
implicitly set to zero everywhere in this paper.

Let us now mention some topological subtleties. First of all, the
expression (\ref{a1}) is, strictly speaking, only well-defined
locally \cite{Cheung:1997}. Secondly, the above proof of
(\ref{singular}) was essentially local, a point directly linked to
the previous one. A way out of this problem is to consider the
following ``framing" regularization procedure (see
\cite{Jaroszewicz:1990} pp. 284-285, and references therein) :
Let
$P({\cal M}_p)\wedge P({\cal M}_p)$ be a product of Poincar\'e
duals taken at the same point. We replace it by the non-singular
product
$P({\cal M}_p)\wedge P({\cal M}_p^\epsilon)$, where ${\cal
M}_p^\epsilon$ is a manifold (i) ``close" to ${\cal M}_p$ the
maximal separation of which is measured by
$\epsilon$, and (ii) without any intersection with ${\cal
M}_p$ : ${\cal M}_p\bigcap{\cal M}_p^\epsilon =\emptyset$.
Eventually, one takes the limit of vanishing $\epsilon$.

This framing procedure should also be applied to the action itself
for every appearing singular product. For instance the cubic CS
term
$dA\wedge dA\wedge A$ also needs to be regularized since
$A$ contains a delta-like singularity. Therefore, three non--intersecting
branes are required to regularize the CS action: the original
Dirac brane worldsheet, ${\cal M}_2$, and two (auxiliary)
displaced worldsheets, ${\cal M}_2^\epsilon$ and
${\cal M}_2^{\epsilon '}$ ($\epsilon\rightarrow 0$, $\epsilon'\rightarrow 0$).

\end{document}